# From Chemically to Physically Induced Pluripotency in Stem Cell

Liaofu Luo

Department of Physics, Inner Mongolia University, Hohhot, 010021 China

**Abstract**

A quantum model on the chemically and physically induced pluripotency in stem cells is proposed. Based on the conformational Hamiltonian and the idea of slow variables (molecular torsions) slaving fast ones the conversion from the differentiate state to pluripotent state is defined as the quantum transition between conformational states. The transitional rate is calculated and an analytical form for the rate formulas is deduced. Then the dependence of the rate on the number of torsion angles of the gene and the magnitude of the rate can be estimated by comparison with protein folding. The reaction equations of the conformational change of the pluripotency genes in chemical reprogramming are given. The characteristic time of the chemical reprogramming is calculated and the result is consistent with experiments. The dependence of the transition rate on physical factors such as temperature, PH value and the volume and shape of the coherent domain is analyzed from the rate equation. It is suggested that by decreasing the coherence degree of some pluripotency genes a more effective approach to the physically induced pluripotency can be made.

## Introduction

Induced pluripotent stem cells (iPSC) were firstly generated by Yamanaka in 2006. They isolated four key pluripotency genes Oct-3/4, SOX2, c-Myc and Klf4 essential for the production of pluripotent stem cells and successfully transformed human fibroblasts into pluripotent stem cells with a retroviral system [1]. The genomic integration of the transcription factors limits the utility of this approach because of the risk of mutations being inserted into the target cell's genome. However, recently Deng et al reported in July 2013 that iPSC can be created chemically without any gene modification [2]. They used a cocktail of seven small-molecule compounds to induce the mouse somatic cells into stem cells (which they called Chemically iPSC or CiPSC) with a higher efficiency up to 0.2%. On the other hand, Su et al reported iPSC can be created directly through a physical approach [3][4]. They indicated that the sphere morphology helps maintaining the stemness of stem cells and proved that, due to the forced growth of cells on low attachment surface the neural progenitor cells can be generated from fibroblasts directly without introducing exogenous reprogramming factors (we call it Physically iPSC or PiPSC). The stimulus-triggered acquisition of pluripotency was proposed tentatively and retracted soon [5,6]. More rigorous experiments with larger statistics on the physical effects on stem cells and the physically-induced pluripotency are awaited for. On the other hand, although many achievements in stem cell experiments have been made, from the point of theory, the mechanism for the chemically-physically iPSC is still one of the most puzzling and confusing problems to be understand. We shall give a quantum theory regarding the acquisition of pluripotency and make an estimate on the probability of the conversion rate. The CiPSC reported in [2] will be quantitatively analyzed and calculated in more detail. Based on the discussion of the factors influencing the conversion a new model of physically-induced pluripotency will also be proposed .

The fundamental processes in CiPSC include small molecules CHIR, 616452, and ESK interacting with key pluripotency-related genes Sall4 and Sox2 to enhance their expression in the early phase to activate the chemical reprogramming, and then small molecule DZNep (as a epigenetic modulator) interacting with gene Oct-4 to enhance its expression in the late phase to switch the process[2]. It is reasonable to suppose that when the small-molecule compounds are bound to the pluripotency gene it causes a sudden change in the molecular conformation (or shape) of the gene, namely a leap (quantum transition) from one of the torsional minima to the another of the molecule [7].   The same story runs in PiPSC. In the PiPSC the three-dimensional (3D) sphere formation is the only important factor to promote the reprogramming without involvement of any exogenous genes, RNAs, proteins or even small molecules. It was speculated that 3D sphere cultures may provide a microenvironment to promote cell dedifferentiation or reprogramming [3]. The further studies indicated the overexpression of gene Sox2 in 3D sphere culture plays a key role in the reprogramming event [4]. So it is natural to assume that certain conformational changes of Sox2 and other genes may have happened in the reprogramming. Therefore, both CiPSC and PiPSC can be studied on the same foot by using the theory of molecular conformational change recently proposed by us [7].

## A quantum model on the induced pluripotency

*Model*   Consider the CiPSC system including a key pluripotency-related gene as a macromolecule and some small molecules interacting with it. The PiPSC can be defined as well when the small molecules are switched off. Apart from 6 translational and rotational degrees of freedom the bond lengths, bond angles, torsion (dihedral) angles of the macromolecule and the coordinate of small molecules relative to the gene and the frontier electrons form a complete set of microscopic variables to describe the system. Among these variables the torsions are slow and others are fast. Following Haken's synergetics, the slow variables always slave the fast ones.

Torsion vibration energy is 0.03-0.003 ev, the lowest **in all forms of biological energies**, even lower than the average thermal energy per atom at room temperature (0.04 eV in 25°C); the torsion angles are easily changed even at physiological temperature. Moreover, the torsion motion has two other important peculiarities. First, due to the strong dependence of the Shannon information quantity on oscillator frequency, the torsion vibration may play an important role in the transmission of information in the biological macromolecular system. Second, different from stretching and bending the torsion potential generally has several minima with respect to angle coordinate that correspond to several stable conformations. We have proved that the small asymmetry in potential (which does exist for a real macromolecule) would cause the strong localization of wave functions and the localized quantum conformational state can well be defined for a biological macromolecule. Therefore, the molecular conformation is defined mainly by the torsion coordinate. After conformation defined through torsion, the quantum transition between conformational states can be calculated.

By the adiabatically elimination of fast variables we obtain the Hamiltonian $H'$ describing the conformational transition of the gene. The matrix element of $H'$ between quantum conformational states, namely between differentiation state $|k\rangle$ and pluripotency state $|k'\rangle$, is denoted by $\langle k' | H' | k \rangle$. Then the transition rate $|\langle k' | H' | k \rangle|^2$ can be deduced from the quantum model [7-8].

*Transitional rate for induced pluripotency.* The Hamiltonian of the gene-molecule system can be expressed as

$$H = H_S(q, \frac{\partial}{\partial q}) + H_F(x, \frac{\partial}{\partial x}; q) \tag{1}$$

where $H_N$ is the slow variable Hamiltonian and the slow variables include the torsion angles of the pluripotency-related gene and those of the small molecule interacting with gene (the latter, the number of torsion angles for small molecules is generally small and can be neglected), $H_F$ is the fast variable Hamiltonian and the fast variables include the bond stretching / bending, the electronic variables and the coordinates of the small molecule relative to the gene.

Equation (1) can be solved under the adiabatic approximation,

$$M(q, x) = y(q) j(x, q) \tag{2}$$

and these two factors satisfy

$$H_F(x, \frac{\partial}{\partial x}; q) j_a(x, q) = e^a(q) j_a(x, q) \tag{3}$$

$$\{H_S(q, \frac{\partial}{\partial q}) + e^a(q)\} y_{kna}(q) = E_{kna} y_{kna}(q) \tag{4}$$

respectively. Here $a$ denotes the quantum state of fast variables, and $(k, n)$ refer to the quantum numbers of torsional conformation and torsional vibration of the gene-molecule system.

Because Eq (4) is not a rigorous eigenstate of Hamiltonian $H_S + H_F$, there exist transitions between adiabatic states that result from the off–diagonal elements [9]

$$\int M^+_{k'n'a'}(H_S + H_F) M_{kna} dq dx = E_{kna} d_{kk'} d_{nn'} d_{aa'} + \langle k'n'a' | H' | kna \rangle \tag{5}$$

$$\langle k'n'a' | H' | kna \rangle = \int y^+_{k'n'a'}(q) \{-\sum_j \frac{\hbar^2}{2I_j} \int j^+_{a'} (\frac{\partial^2 j_a}{\partial q_j^2} + 2 \frac{\partial j_a}{\partial q_j} \frac{\partial}{\partial q_j}) dx\} y_{kna}(q) dq \tag{6}$$

Here $H'$ is a Hamiltonian describing the conformational transition of the gene between the quantum conformational state $|kna\rangle$ (or simply denoted as $|k\rangle$) and

$|k'n'a'\rangle$ (or $|k'\rangle$) and $I_j$ is the inertial moment of the $j$-th torsion mode.

Based on the $q$ dependence of fast-variable wave function $j_a(x, q)$ deduced by the perturbation method the matrix element $\langle k'n'a' | H' | kna \rangle$ can be obtained. Through tedious calculations [7][8] one obtains the differentiate-to-pluripotent conformational transition rate [7]

$$W = \frac{2p}{\mathbf{h}^2 \bar{w}'} I_V' I_E' \tag{7}$$

$$I_V' = \frac{\mathbf{h}}{\sqrt{2p}dq} \exp\{\frac{\Delta G}{2k_B T}\} \exp\{\frac{-(\Delta G)^2}{2\bar{w}^2 (dq)^2 k_B T \sum_j^N I_j}\} (k_B T)^{1/2} (\sum_j^N I_j)^{-1/2} \tag{8}$$

$$I_E' = \sum_j^M |a_{a'a}^{(j)}|^2 \cong M\bar{a}^2 \tag{9}$$

where $I_V'$ is slow-variable factor and $I_E'$ fast-variable factor, $\Delta G$ is the free energy decrease per molecule between initial and final states, $N$ is the number of torsion modes participating in the quantum transition coherently, $I_j$ denotes the inertial moment of the atomic group of the j-th torsion mode, $\bar{w}$ and $\bar{w}'$ are the torsion potential parameters $w_j$ and $w_j'$ averaged over $N$ torsion modes in initial and final state, respectively, $dq$ is the averaged angular shift between initial and final torsion potential, $M$ is the number of torsion angles correlated to fast variables, $k_B$ is Boltzmann constant and $T$ is absolute temperature.

Eqs (7-9) give the rate of gene conformational transition in CiPSC. As the small-molecule variables not appearing in Eq (1), the same equations deduced above describes the rate of gene conformational transition in PiPSC.

Eqs (7-9) are introduced for calculating the transition from differentiate to pluripotent state. In principle, it holds equally well for the reverse process, from pluripotent to differentiate states. The rate of the reverse process is obtained by the replacement of $\Delta G$ by $-\Delta G$ and $\bar{w}'(\bar{w})$ by $\bar{w}$ ($\bar{w}'$) in $W(k \to k')$. One has

$$\ln\{\frac{W(k \to k')}{W(k' \to k)}\} = \frac{\Delta G}{k_B T} + \frac{(\Delta G)^2}{2k_B T \bar{w}^2 (dq)^2 \sum_j^N I_j} (\frac{\bar{w}^2 - \bar{w}'^2}{\bar{w}'^2}) + \ln\frac{\bar{w}}{\bar{w}'} \cong \frac{\Delta G}{k_B T} \tag{10}$$

We assume differentiation state $k$ has lower conformational energy (in torsion ground-state) while pluripotency state $k'$ has higher conformational energy (in torsion excited-state). Therefore, due to $\Delta G < 0$, the acquisition of pluripotency is a small-probability event as seen from Eq(10).

*More statistical analyses on protein folding and generalization to pluripotency transition*   The rate formula Eqs(7)-(9) are deduced from very general assumptions on molecular conformational change. They were tested on the protein folding problem. The statistical analyses of 65 two-state protein folding given in [10] shows these formula are in good accordance with experimental rates [11]. For a two-state protein the torsion number $N$ is calculated by the numeration of all main-chain and side-chain dihedral angles on the polypeptide chain. Typically $N$ takes 100 – 300 for a typical two-state protein. To apply these formula on the pluripotency conversion in stem cells one should study the conformational change of the gene. For a pluripotency gene the DNA chain contains alternating links of phosphoric and sugar (deoxyribose) and every sugar is attached to a nitrogen base. Following IUB/IUPAC there are 7 torsion angles for each nucleotide, namely

$$a(O3'-P-O5'-C5'),$$
$$b(P-O5'-C5'-C4'),$$
$$g(O5'-C5'-C4'-C3'),$$
$$d(C5'-C4'-C3'-O3'),$$
$$e(C4'-C3'-O3'-P),$$
$$V(C3'-O3'-P-O5')$$

and $c(O4'-C1'-N1-C2)$ (for Pyrimidine) or $c(O4'-C1'-N9-C2)$ (for Purine), of which many have more than one advantageous conformations (potential minima). So DNA molecule contains more torsion degrees of angles than a protein. Although it is no problem from the theoretical point that Eqs(7)-(9) can be used as well for DNA, we should study the torsion number dependence of the conformation-transition rate.

From Eqs(7)-(9) we find the main dependence factors of the rate is the free energy decrease (per molecule between initial and final states) $\Delta G$ and the fast-variable factor $\bar{a}^2$ ( the square of the matrix element of the fast-variable Hamiltonian operator $H_F$ changing with torsion angle averaged over $M$ modes). We shall study how these two factors depend on the torsion number $N$. For protein folding or other macromolecular conformational change not involving chemical reaction and electronic transition the fast variable includes only bond lengths and bond angles of the macromolecule. In this case a simplified expression for the fast-variable factor $I'_E$ can be deduced. Consider the globular protein folding as a typical example. Suppose the coherent transition area of the molecule is a rotational ellipsoid with semi major axis $a$ and semi minor axis $b$, and $j_a(x, q)$ a constant in the ellipsoid. The matrix element of the stretching-bending Hamiltonian is

$$(H_F)_{a'a} \cong \frac{1}{V_{ep}^2} \int j_{a'}^*(\mathbf{r},\mathbf{r}')U j_a(\mathbf{r},\mathbf{r}')d^3\mathbf{r} d^3\mathbf{r}' \tag{11}$$

$$U \approx \sum_{d=12,10,6,1} \frac{c_d}{(|\mathbf{r}-\mathbf{r}'|)^d}.$$

where the form of U is assumed following the general potential function for peptides[11]. It gives

$$(H_F)_{a'a} = \frac{1}{V_{ep}} \sum_d c_d \int_{ep} (|\mathbf{r}-\mathbf{r}'|)^{-d} d^3(\mathbf{r}-\mathbf{r}')$$

$$\cong \frac{1}{V_{sp}} \frac{a^2}{b^2} \sum_{d=12,10,6,1} c_d \int_{sp} d\Omega dr r^2 (r+h)^{-d} \tag{12}$$

($h$ is a cutoff parameter) where $V_{ep} = \frac{4p}{3}ab^2$ is the volume of the ellipsoid, $V_{ep} = \frac{4p}{3}a^3$ the volume of a sphere, the first integral is taken over the ellipsoid and the second integral over the sphere. Because $V_{sp}$ and $M$ is proportional to $N$, and the effective coupling $\frac{\partial c_d}{\partial q_j}$ is inversely proportional to the assumed interacting-pair number (about $N^2$), while the weak dependence of the integral $\int_{sp} d\Omega dr r^2 (r+h)^{-d}$ on the sphere radius can be neglected, from Eq (12) we estimate

$$M\bar{a}^2 = cfN^{-5} \quad (f = \frac{a^4}{b^4}) \tag{13}$$

where $f$ is a shape parameter. It means the fast-variable factor $I_E'$ is inversely proportional to $N^5$. This can be understood by only a small fraction of interacting-pairs (about one in $N^2$) correlated to $q_j (j=1,...,N)$ in $\frac{\partial H_F}{\partial q_j}$. It is reasonable to assume the similar relation holds for the system of nucleotides. With $\sum_j^N I_j \cong NI_0$ and Eq (13) inserted into Eq (7) to (9) we obtain the rate

$$W = \frac{\sqrt{2p}}{\mathbf{h} dq \bar{w}'} (k_B T)^{1/2} \exp\{\frac{\Delta G}{2k_B T}\}(NI_0)^{-1/2} \exp\{\frac{-(\Delta G)^2}{2\bar{w}^2 (dq)^2 k_B T NI_0}\}(cfN^{-5}) \tag{14}$$

or the relation of logarithm rate with respect to $N$ and $\Delta G$

$$\ln W = \frac{\Delta G}{2k_B T} - \frac{(\Delta G / k_B T)^2}{2rN} - 5.5 \ln N + \ln c_0 f \qquad (15)$$

where

$r = I_0 \bar{w}^2 (dq)^2 / (k_B T)$ is a torsion energy-related parameter and

$c_0 = \frac{\sqrt{2p}}{\mathbf{h} dq} (\frac{k_B T}{\bar{w}'^2 I_0})^{1/2} c$ is an $N$-independent constant.

The relation of $\ln W$ with $N$ given by Eq (15) is in good accordance with the statistical analyses of 65 two-state protein folding rates. The correlation coefficient $R$ between experimental logarithm folding rate and theoretical $\ln W$ is 0.783. [7,11].

To find the relation between free energy $\Delta G$ and torsion number $N$ we consider the statistical relation of $\frac{\Delta G}{2k_B T} - \frac{(\Delta G)^2}{2(k_B T)^2 rN}$ that occurs in rate equation (8) or (15). Set

$$\frac{\Delta G}{2k_B T} - \frac{(\Delta G)^2}{2(k_B T)^2 rN} = y, \qquad -\frac{1}{N} = x \qquad (16)$$

The linear regression between y and x is shown in Fig 1 for 60 two-state proteins.

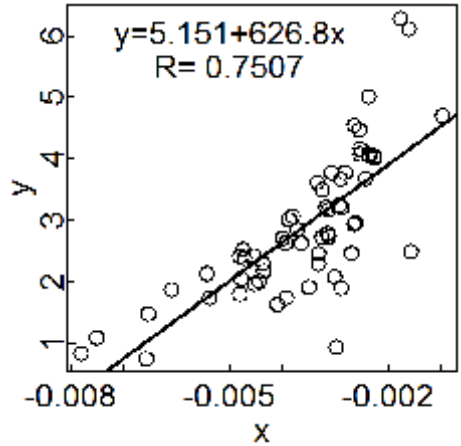

Fig 1  Statistical relation for 60 two-state proteins. Experimental data are taken from 65- protein set [10]. Five proteins denatured by temperature in the set have been omitted in our statistics.

The correlation $R$ equals 0.757. In the statistics of Fig 1 the parameter $r = 0.097$ has been assumed. If the variation of $r$ of different proteins in the set is considered then $R$ will be increased. In virtue of Eqs (15) (16) we obtain an approximate expression for transitional rate $\ln W$ versus $N$ for protein folding

$$\ln W = A_{pt} - \frac{B_{pt}}{N} - 5.5\ln N + c_{prot} \qquad (17)$$

$$A_{pt}=5.151, \quad B_{pt}=626.8$$

$c_{prot}$ is an N-independent constant. It gives $W$ increasing with $N$, attaining the maximum at $N_{max}=B/5.5$, then decreasing with power law $N^{-5.5}$. For the reverse process (protein unfolding), Eq (17) still holds but $A_{pt}=-5.151$.

The above equation (17) can be generalized to pluripotency transition in DNA. The rate is given by

$$\ln W_a = A_a - \frac{B_a}{N} - 5.5\ln N + c_{DNA} \qquad (18)$$

$$A_a = -5.151, \quad B_a = 626.8$$

from differentiate (torsion-ground) to pluripotent (torsion excited) state and

$$A_a = 5.151, \quad B_a = 626.8$$

from pluripotent (torsion excited) to differentiate (torsion-ground) state, $c_{DNA}$ is an N-independent constant.

Eqs (7) to (9) are the fundamental equations for quantum transition in CiPSC and PiPSC. The key points in the above deduction lie in the definition of the differentiate and pluripotent state in terms of conformational state and the application of adiabatic approximation (slaving principle) to deduce the general transition formula between them. Eq(18) can be used for the estimate of how the rate changes with $N$ in different processes. The important thing is: In spite of the various complex pathways in the cell's development both CiPSC and PiPSC have the same quantum conformational transition as a key step in the reprogramming of stem cells. Moreover, the stability of quantum state gives a natural explanation on the maintenance of the stemness of stem cells.

## Chemically induced pluripotency discussed in quantum model

Pluripotent stem cells were successfully induced from mouse somatic cells by small-molecule compounds [2]. The schematic diagram (Fig 2) taken from [2] illustrates the stepwise establishment of the pluripotency circuity during chemical reprogramming. There are seven pluripotency genes, namely Oct4, Sox2, Nanog, Sall4, Sox17, Gata4 and Gata6, involved in the circuitry.

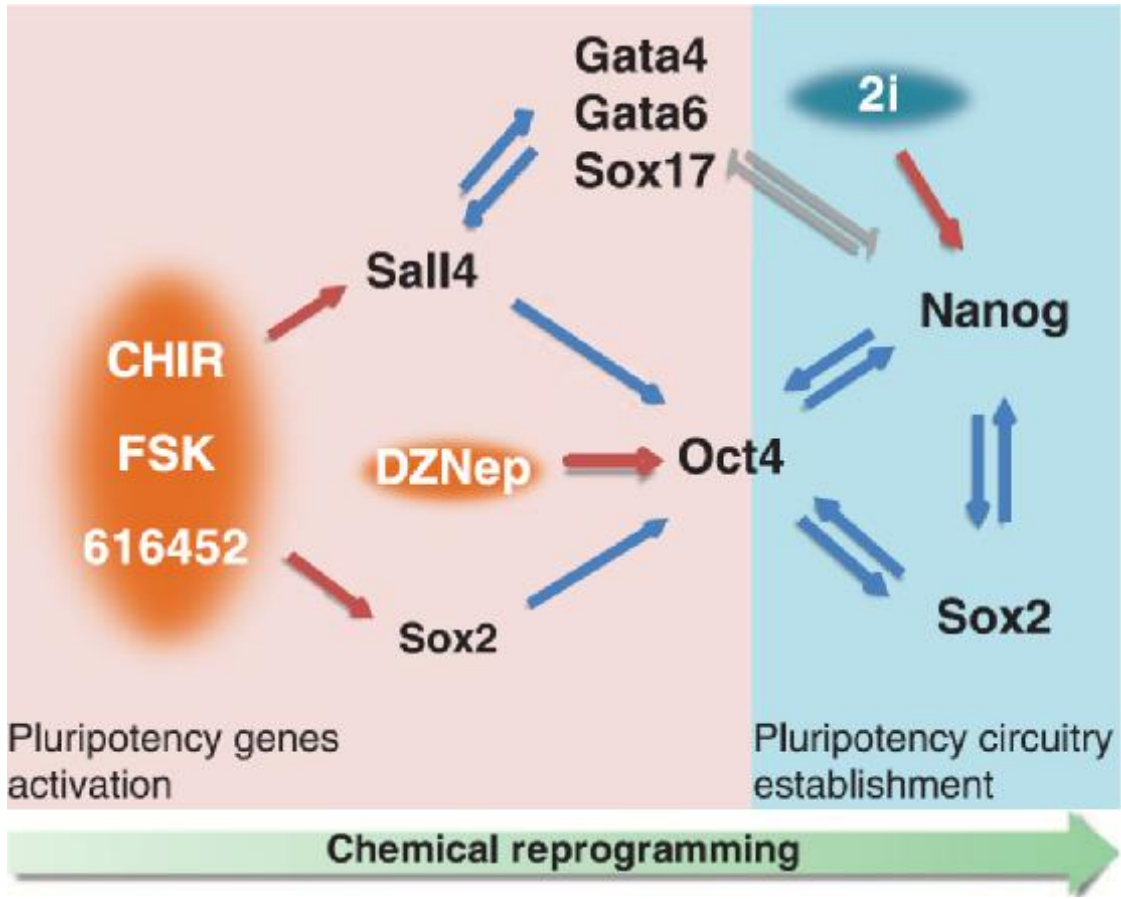

Fig 2 The diagram is taken from literature [2]. The schematic diagram illustrates the pluripotency circuity during chemical reprogramming. CHIR(C), FSK(F), 616452(6) and DZNep(Z) are small molecules. Oct4, Sox2, Nanog, Sall4，Sox17, Gata4 and Gata6 are seven genes involved in the pluripotency circuitry establishment.

Assume only the torsion transition of single gene is considered. For double-directional interaction of gene $A$ and $B$ in Fig 2, we assume a conformational transition from torsion-ground to torsion-exited state of gene A or its reverse in companion of gene B,

$$A + B \rightarrow (or \leftarrow) A^* + B,$$

or the transition of gene B in companion of gene A,

$$A + B \rightarrow (or \leftarrow) A + B^*$$

where star * after a gene means its torsion-excited state. Following Fig 2 the reaction equations of the conformational change of seven genes are written below. The first 3 equations describe the conformational transition of pluripotency genes under the action of small molecules CHIR(C), FSK(F), 616452(6) and DZNep(Z). The next 7 equations describe the conformational transition under gene interaction.

1) C6F+Sox2 $\xrightarrow{k_1}$ Sox2*+ C6F;

2) C6F+Sall4 $\xrightarrow{k_2}$ Sall4*+ C6F;

3) Sox2+Sall4+Oct4+Z $\xrightarrow{k_3}$ Sox2+Sall4+Oct4*+Z （companions Sox2 and Sall4 can be replaced by their excited states）；

4) Sall4+Gata4 $\xrightarrow{k_4}(or \xleftarrow{k_4'})$ Sall4*+Gata4 (companion Gata4 can be replaced by Gata6，Sox17 or their excited states )

5) Gata4+ Sall4 $\xrightarrow{k_5}(or \xleftarrow{k_5'})$ Gata4*+Sall4 (companion Sall4 can be replaced by Nanog or their excited states)

6) Gata6+ Sall4 $\xrightarrow{k_6}(or \xleftarrow{k_6'})$ Gata6*+Sall4 (companion Sall4 can be replaced by Nanog or their excited states)

7) Sox17+ Sall4 $\xrightarrow{k_7}(or \xleftarrow{k_7'})$ Sox17*+Sall4 (companion Sall4 can be replaced by Nanog or their excited states )

8) Nanog+Gata4 $\xrightarrow{k_8}(or \xleftarrow{k_8'})$ Nanog*+Gata4（companion Gata 4 can be replaced by Gata6，Sox17 or their excited states in 2i-medium，by Oct4，Sox2 or their excited states without 2i-medium）

9) Oct4+Nanog $\xrightarrow{k_9}(or \xleftarrow{k_9'})$ Oct4*+Nanog   (companion Nanog can be replaced by Sox2 or their excited states )

10）Sox2+ Nanog $\xrightarrow{k_{10}}(or \xleftarrow{k_{10}'})$ Sox2*+ Nanog  (companion Nanog can be replaced by Oct4 or their excited states )

The pluripotency circuitry is divided into three sub-circuits: circuit *a* of small-molecule reaction 1 to 3; circuit *b,* including reaction 4 to 8 from Sall4 to Gata 4 (Gata6,Sox17) to Nanog then in the reverse direction to Sall4 as seen from Fig 2; and circuit *c,* including reaction 8 to 10 from Nanog to Oct4 to Sox 2 then in the reverse direction to Nanog.

Denote the dimensionless concentration of these molecules by x(Oct4), y(Sox2), z( Nanog）, w( Sall4), s(Sox17), u(Gata4）and v( Gata6）respectively, and the concentration of their torsion-excited state by x'(Oct4*), y'(Sox2*), z'( Nanog*）, w'( Sall4*), s'(Sox17*), u'(Gata4*) and v'( Gata6*）respectively.  Denote the dimensionless concentration of small molecules C6F and DZNep by *a* and *b* respectively and both are assumed as constant in the reaction. The torsion excitation of these small molecules are neglected.  The reaction equations for pluripotency gene concentrations are

$$\frac{dx}{dt} = -k_3 bx(y+y')(w+w') - k_9 x(y+y'+z+z') + k_9' x'(y+y'+z+z') = -\frac{dx'}{dt}$$
(19.1)

$$\frac{dy}{dt} = -k_1 ay - k_{10} y(z+z'+x+x') + k_{10}' y'(z+z'+x+x') = -\frac{dy'}{dt} \qquad (19.2)$$

$$\frac{dz}{dt} = -k_8(u+u'+v+v'+s+s'+x+x'+y+y')z + k_8'(u+u'+v+v'+s+s'+x+x'+y+y')z'$$

$$= -\frac{dz'}{dt} \qquad (19.3)$$

$$\frac{dw}{dt} = -k_2 aw - k_4(u+u'+v+v'+s+s')w + k_4'(u+u'+v+v'+s+s')w' = -\frac{dw'}{dt}$$
(19.4)

$$\frac{du}{dt} = -k_5(w+w'+z+z')u + k_5'(w+w'+z+z')u' = -\frac{du'}{dt} \qquad (19.5)$$

$$\frac{dv}{dt} = -k_6(w+w'+z+z')u + k_6'(w+w'+z+z')u' = -\frac{dv'}{dt} \qquad (19.6)$$

$$\frac{ds}{dt} = -k_7(w+w'+z+z')s + k_7'(w+w'+z+z')s' = -\frac{ds'}{dt} \qquad (19.7)$$

Set

$$x+x'=c_1, \quad y+y'=c_2, \quad z+z'=c_3,$$
$$w+w'=c_4, \quad u+u'=c_5, \quad v+v'=c_6, \quad s+s'=c_7 \qquad (20)$$

One obtains the steady states immediately

$$x_0 = \frac{k_9' c_1 (c_2+c_3)}{(k_9+k_9')(c_2+c_3)+k_3 b c_2 c_4} \quad \text{(for Oct4)}$$

$$y_0 = \frac{k_{10}' c_2 (c_1+c_3)}{(k_{10}+k_{10}')(c_1+c_3)+k_1 a} \quad \text{(for Sox2)}$$

$$z_0 = \frac{k_8' c_3}{(k_8+k_8')} \quad \text{(for Nanog)}$$

$$w_0 = \frac{k_4' c_4 (c_5+c_6+c_7)}{(k_4+k_4')(c_5+c_6+c_7)+k_2 a} \quad \text{(for Sall4)}$$

$$u_0 = \frac{k_5' c_5}{(k_5+k_5')} \quad \text{(for Gata4);}$$

$$v_0 = \frac{k_6' c_6}{(k_6+k_6')} \quad \text{(for Gata6)}$$

$$s_0 = \frac{k_7' c_7}{(k_7+k_7')} \quad \text{(for Sox17)} \qquad (21)$$

and proves the stability of these steady states. Moreover, the relaxation times for attaining the steady state are obtain

$$t_1 = 1/\{(k_9 + k_9')(c_2 + c_3) + k_3 b c_2 c_4\} \quad \text{(for Oct4)}$$

$$t_2 = 1/\{(k_{10} + k_{10}')(c_1 + c_3) + k_1 a\} \quad \text{(for Sox4)}$$

$$t_3 = 1/\{(k_8 + k_8')(c_1 + c_2 + c_5 + c_6 + c_7)\} \quad \text{(for Nanog)}$$

$$t_4 = 1/\{(k_4 + k_4')(c_5 + c_6 + c_7) + k_2 a\} \quad \text{(for Sall4)}$$

$$t_5 = 1/\{(k_5 + k_5')(c_3 + c_4)\} \quad \text{(for Gata4)}$$

$$t_6 = 1/\{(k_6 + k_6')(c_3 + c_4)\} \quad \text{(for Gata6)}$$

$$t_7 = 1/\{(k_7 + k_7')(c_3 + c_4)\} \quad \text{(for Sox17)} \quad (22)$$

From the rates $\{k_i\}$ of ten torsion transitions one can define three characteristic times corresponding to three sub-circuits,

$$t_a = \frac{1}{k_1} + \frac{1}{k_2} + \frac{1}{k_3} \quad (23.1)$$

$$t_b = \frac{1}{k_4} + \frac{1}{\bar{k}} + \frac{1}{k_8} + \frac{1}{k_4'} + \frac{1}{\bar{k}'} + \frac{1}{k_8'}$$

$$\frac{1}{\bar{k}} = \frac{1}{3}(\frac{1}{k_5} + \frac{1}{k_6} + \frac{1}{k_7}), \quad \frac{1}{\bar{k}'} = \frac{1}{3}(\frac{1}{k_5'} + \frac{1}{k_6'} + \frac{1}{k_7'}) \quad (23.2)$$

$$t_c = \frac{1}{k_8} + \frac{1}{k_9} + \frac{1}{k_{10}} + \frac{1}{k_8'} + \frac{1}{k_9'} + \frac{1}{k_{10}'} \quad (23.3)$$

To calculate these characteristic times and relaxation times we estimate the rates $k_i$ and $k'_i$ by using eq (18). By setting $k_\alpha = k_1$, $k_2$, $k_3$, $k_4$, $k_4'$ … … $k_{10}$ or $k_{10}'$ one calculates the ratio of the transition rate to the folding rate $k_{pt}$ of a typical two-state

protein (with $N=N_{pt}$, $1/N_{pt} = 0.003$, $k_{pt} = 10^4 s^{-1}$) as

$$\ln\frac{k_a}{k_{pt}} = (A_a - A_{pt}) - (\frac{B_a}{N_a} - \frac{B_{pt}}{N_{pt}}) - 5.5\ln\frac{N_a}{N_{pt}} + (c_{DNA} - c_{pt}) \qquad (24)$$

where $N_a$ is the torsion number of the α-th transition. Assuming the torsion number $N_a$ is proportional to the DNA sequence length of the related gene, $N_a = q \times$ (nucleotide number in α-th DNA), one has [12]

$$N_{Oct4}=6357q, \quad N_{Sox2}=2512q, \quad N_{Nanog}=6661q, \quad N_{Sall4}=18466q,$$

$$N_{Gata4}=2952q, \quad N_{Gata6}=55793q, \quad N_{Sox17}=32812q \qquad (25)$$

By using Eq (24) (25) we obtain

$$\ln\frac{k_1}{k_{pt}} = \ln\frac{k_{10}}{k_{pt}} = \frac{B}{N_{pt}} - \frac{B}{N_{Sox2}} - 5.5\ln\frac{N_{Sox2}}{N_{pt}}$$

$$\ln\frac{k_2}{k_{pt}} = \ln\frac{k_4}{k_{pt}} = \frac{B}{N_{pt}} - \frac{B}{N_{Sall4}} - 5.5\ln\frac{N_{Sall4}}{N_{pt}}$$

$$\ln\frac{k_3}{k_{pt}} = \ln\frac{k_9}{k_{pt}} = \frac{B}{N_{pt}} - \frac{B}{N_{Oct4}} - 5.5\ln\frac{N_{Oct4}}{N_{pt}} \qquad \text{etc} \qquad (26)$$

and

$$\ln\frac{k_4}{k_4'} = \ln\frac{k_5}{k_5'} = ... = \ln\frac{k_{10}}{k_{10}'} = -2A_{pt} = -10.3 \qquad (27)$$

In the calculation of Eq (26) the parameters $c_{DNA} - c_{prot} = 10.3$ and q=6 or 5 are taken. The numerical results are given in Table 1.

**Table 1   Rates of single torsion transition in related gene**

| Rate | $k_1$ | $k_2$ | $k_3$ | $k_4$ | $k_5$ | $k_6$ | $k_7$ | $k_8$ | $k_9$ | $k_{10}$ |
|---|---|---|---|---|---|---|---|---|---|---|
| Gene | Sox2 | Sall4 | Oct4 | Sall4 | Gata4 | Gata6 | Sox17 | Nanog | Oct4 | Sox2 |
| N/q | 2512 | 18466 | 6357 | 18466 | 2952 | 55793 | 32812 | 6661 | 6357 | 2512 |
| $\ln \dfrac{k_i}{k_{pt}}$ (q=6) | -19.08 | -30.05 | -24.19 | -30.05 | -19.97 | -36.14 | -33.22 | -24.45 | -24.19 | -19.08 |
| $\ln \dfrac{k_i}{k_{pt}}$ (q=5) | -18.08 | -29.05 | -23.19 | -29.05 | -18.97 | -35.14 | -32.22 | -23.45 | -23.19 | -18.08 |
| Sub-circuit | a | a | a | b | b | b | b | b | c | c |
|  |  |  |  |  |  |  |  | c |  | c |

The eighth reaction of Nanog torsion transition in companion with Gata4, Gata6 and Sox17 belongs to sub-circuit $b$, and that in companion with Oct 4 and Sox 2 belongs to sub-circuit $c$.

It leads to

$$t_a \approx \frac{1}{k_2} = e^{30.05} t_{pt} = 26d / 0.2\% (q=6), \quad 9.5d / 0.2\% (q=5) \tag{28.1}$$

$$t_b \approx \frac{1}{k} \gg t_a \quad \text{(without 2i-medium)} \tag{28.2}$$

$$t_c \approx \frac{1}{k_8} + \frac{1}{k_9} = e^{24.19} t_{pt} + e^{24.45} t_{pt} = 85d (q=6), \quad 31d (q=5) \tag{28.3}$$

The relaxation times, Eq(22), depends not only on rate $k_i$ but also on concentration $c_i$.

The characteristic time $t_a$ (28.1) is basically in consistency with the experimental data of pluripotent stem cells generated at a frequency up to 0.2% on day 30-40 [2] as we notice that more time is needed for post-transitional steps after the quantum transition to acquire the fully reprogrammed cell.  On the other hand we notice that only after switching to 2i-medium the transitions in cicuit $b$ speed up and $t_b < t_a$. As $t_b$ is neglected the total pluripotency transition time $t$ equals

$$t = t_a + t_b + t_c = t_a + t_c \approx t_a \tag{29}$$

consistent with experiment.

Remarks

1   The above calculation shows the pluripotency transition time is in the order of several days to weeks, much longer than the protein folding time. The reason is the coherence degree $N$ in DNA torsion transition much larger than that in protein folding.

2. The transition time $t_b$ is about 150 $t_a$ in absence of 2i medium. The time needed for full reprogramming would be too long if $t_b$ has been taken into account. However, introduce of 2i medium in circuit $b$ can effectively shorten the total pluripotency transition time. Notice that the 2i medium introduced in reaction step 8 only influences the single torsion transition of Nanog gene in companion of Gata 4, Gata6 and Sox17, but has nothing to do with the same transition of Nanog in companion of Oct4 and Sox2 in circuit $c$.

3. In circuit $b$ and circuit $c$ the single torsion transition is in double directions. The reverse transition from torsion-excited to ground state is much faster (Eq 27), providing a positive feedback mechanism to the network.

4. Work [2] introduced small-molecule interaction to activate the pluripotency genes. It is expected that the pluripotency transition may be speeded up by improving the small-molecule interaction. However, the present theory gives a lower limit for the pluripotency transition in CiPSC, namely $t_{\text{limit}} = t_c \cong \frac{1}{153} t_a$.

5. Above points 1-4 are deduced irrespective of parameter choice. The only adjustable parameters in the theory are $c_{\text{DNA}}$ and q but they occur in the combination of $c_{\text{DNA}}$-5.5lnq. The parameter q means the average number of torsions in each nucleotide. The upper limit of q is 7 for a single chain, so the choice of q=6 or 5 seems reasonable because a part of torsional potentials have only one minimum. The parameter $c_{\text{DNA}}$ is related to the free energy $\Delta G$ in the transition. We have assumed $c_{DNA} - c_{prot} = 2A_{pt}$ which is in the correct range of free energy change. Of course, one may use alternative parameter choices but remains $c_{\text{DNA}}$-5.5lnq unaltered to obtain the same result.

6. We have discussed the rate of the acquisition of pluripotency from the quantum transition theory. The transition rate calculated above is for single torsion transition only. In reality the pluripotency can be acquired in multi-torsion transitions, for example, by the equation of gene interaction $A + B \xrightarrow{k} (or \xleftarrow{k'}) A* + B*$. However the rate of multi-torsion transitions is much lower than the single ones and they can always be neglected.

## Physically-induced pluripotency inferred from the quantum theory

We shall study the physical factors influencing the rate of the acquisition of pluripotency from the quantum transition theory. The important dependence factors of the rate inferred from the theory are:

## 1. Transition rate depends on temperature

Assuming the free energy decrease $\Delta G$ in pluripotency genes is linearly dependent of temperature $T$ as in protein folding we obtain the temperature dependence of the transition rate from Eq 7 to 9,

$$\ln W(T) = \frac{S}{T} - RT + \frac{1}{2}\ln T + const. \tag{30}$$

It means the non-Arrhenius behavior of the rate–temperature relationships. The relation was proved in good agreement with the experimental data of protein folding [7][11]. For a typical two-state protein the folding rate changes (increases or decreases) 2- to 7-fold in a temperature range of 40 degrees. Now we predict the same relation also holds for the reprogramming transition for the pluripotency gene. .

## 2. Transition rate depends on PH

Following the biochemical principle the free energy change $\Delta G$ of a reaction is linearly dependent on the logarithm of ion concentration [H$^+$]. One has

$$\frac{\Delta G}{k_B T} = \frac{(\Delta G)_0}{k_B T} + 2.3 b (7 - \text{PH}) \tag{31}$$

where $(\Delta G)_0$ is the standard free energy change at PH 7, $b = O(1)$. Inserting it into Eq (7-9) we obtain

$$\frac{d}{d\text{PH}} \ln W(\text{PH}) < 0 \quad (\text{for} \quad \Delta G < 0) \tag{32}$$

When PH decreases from 7 to 6 (or to 5, to 4, ...) the logarithm rate increases by a quantity

$$d \ln W = 1.15 + \frac{2.3}{rN}(\frac{\Delta G}{k_B T})_0 \cong 1.15 \tag{33}$$

(or $\cong$ 2.3, 3.45, ...) or the rate increases by a factor 3.16 (or 9.97, 31.5, ...). It means one may observe the acidity-induced pluripotency by soaking the tissues in acidic medium below PH 6.0.

## 3. Transition rate depends on volume and shape of the coherent domain

As seen from Eq (18) the transition rate is strongly dependent of the coherence degree $N$ of the gene. One may assume that the coherence degree is proportional to the volume $V$ of the coherent domain in the gene. Therefore, from Eq (18) one has

$$W(V) \approx fV^{-5.5} \exp(-\frac{a}{V}) \tag{34}$$

($a$>0 is a $V$-independent constant). Eq (34) means $W(V)$ grows as $V$ decreases (in the range $V > \frac{a}{5.5}$). To lower down the coherence volume of the pluripotency gene a simple method is trituration of the sample under appropriate constraint.

The protein folding rate is dependent of the shape of the protein. The folding rate for a two-state protein having a quite oblong or oblate ellipsoid shape is several tenfold higher than a spheroid shape [7][10]. This can be seen by the shape parameter $f$ appearing in Eq (13). The shape parameter is another important factor to measure the coherence degree and influence the transition rate. As compared with spheroid the sphere morphology will benefit taking a larger coherence degree N and therefore a smaller transition rate. On the other hand, Le Chatelier's principle states that if a stress is applied to a reaction at equilibrium the equilibrium will be displaced in the direction that relieves the stress. The shape of the cell as a stress applied to the microenvironment of the pluripotency gene will influence the equilibrium of the cell [3][4].

Based on the above analyses we propose a physically induced pluripotency circuitry as shown in Fig 3. The main pluripotency circuitry is established through the cycle of Oct4-Sox2-Nanog and the latter is connected to pluripotency genes X and Y. The genes X and Y are activated through their torsion transitions under the action of physical factors. The reaction equations are

(physical factor)+X $\xrightarrow{k_X}$ X*

(physical factor)+Y $\xrightarrow{k_Y}$ Y*

Nanog+Sox2 $\xrightarrow{k_8}(or \xleftarrow{k_8'})$ Nanog*+Sox2 (companion Sox2 can be replaced by Oct4 or their excited states, or Y*)

Oct4+Nanog $\xrightarrow{k_9}(or \xleftarrow{k_9'})$ Oct4*+Nanog (companion Nanog can be replaced by Sox2 or their excited states, or $X^*$)

Sox2+ Nanog $\xrightarrow{k_{10}}(or \xleftarrow{k_{10}'})$ Sox2*+ Nanog (companion Nanog can be replaced by Oct4 or their excited states)

As an example, suppose X=Sall4 and Y=Sox17. Assume the volume and shape of coherent domain of both genes are changed under physical action. As a result, both the torsion numbers of Sall4 and Sox17 decrease a factor $\lambda$. The total pluripotency transition time of the assumed circuitry is

$$t = t_{XY} + t_c \qquad (35.1)$$

where $t_c$ is given by Eq(28.3) and

$$t_{XY} = \frac{1}{k_X} + \frac{1}{k_Y}$$

$$\ln\frac{k_X}{k_{pt}} = \frac{B}{N_{pt}} - \frac{B}{l\,N_{Sall4}} - 5.5\ln\frac{l\,N_{Sall4}}{N_{pt}}$$

$$\ln\frac{k_Y}{k_{pt}} = \frac{B}{N_{pt}} - \frac{B}{l\,N_{Sox17}} - 5.5\ln\frac{l\,N_{Sox17}}{N_{pt}} \qquad (35.2)$$

The smaller the factor $\lambda$ is the larger the rates $k_X$ and $k_Y$ will be. When $l < l_{threshold} = \frac{1}{4.5}$ one has $t_{XY} < t_c$ and the pluripotency transition time is dominated by $t_c$. If only X is activated by physical factor (decreasing of $N$) then the threshold $l_{threshold} = \frac{1}{2.5}$. The above results are irrespective of q choice.

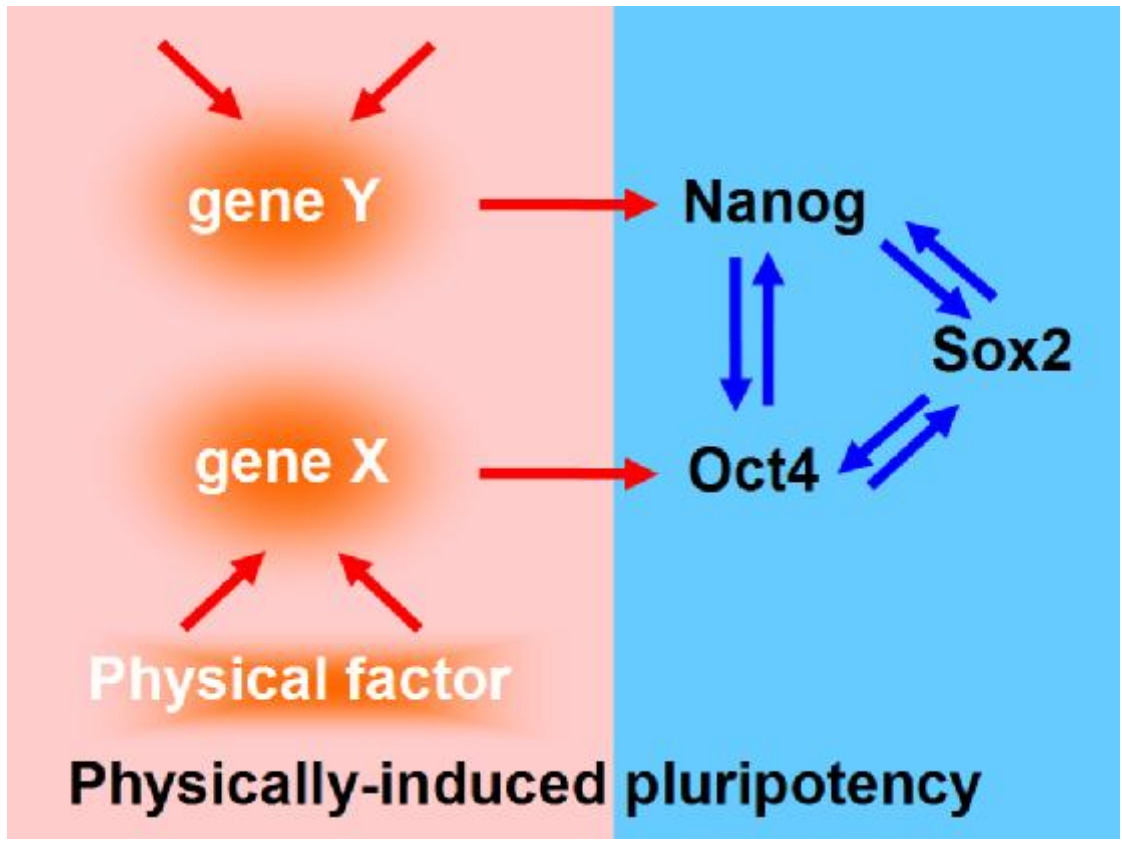

Fig 3  The schematic diagram illustrates the physically-induced pluripotency. The physical factor lowers down the coherence degrees in the torsion transition of the gene X and Y and then establishes the pluripotency conversion in the cycle of Oct4-Sox2-Nanog .

To conclude, we have shown the physical factors such as temperature and acidity can change the rate of torsion transition. So, low PH exposure and heat shock may have observable effect in the pluripotency conversion. However, we found the most effective approach to increase the torsion transition rate is to lower down the coherence degrees of the gene. Based on the proposed model the physically induced pluripotency transition rate can, in principle, increase more than 150 times as compared with that in the known small-molecule induction approach.

Acknowledgement   The author thanks Dr Jun Lu and Dr Judong Zhao for their statistical analyses on protein folding, Dr Yennie Cao and her group for data collection in pluripotency circuity and figure drawing.   He also thanks Dr Yulai Bao for many helps in literature searching.